\documentclass[a4paper,12pt]{article}
\usepackage[T1]{fontenc}
\usepackage[utf8]{inputenc}
\usepackage{lmodern}
\usepackage[left=2.75cm,right=2.65cm,top=2.8cm,bottom=2.6cm]{geometry}
\usepackage{amsmath,amssymb,amsthm,mathtools}
\usepackage{stmaryrd}
\usepackage{array,longtable,booktabs}
\usepackage{enumitem}
\usepackage{microtype}
\usepackage{setspace}
\usepackage{indentfirst}
\makeatletter
\renewcommand\section{\@startsection{section}{1}{\z@}{-3.5ex \@plus -1ex \@minus -.2ex}{2.3ex \@plus .2ex}{\centering\normalfont\Large\bfseries}}
\renewcommand\subsection{\@startsection{subsection}{2}{\z@}{-3.25ex\@plus -1ex \@minus -.2ex}{1.5ex \@plus .2ex}{\normalfont\large\bfseries}}
\makeatother
\begin{document}
\begin{center}\LARGE\bfseries Dynamic Preference Logic of Preference Lifting\end{center}
\begin{center}Xiaochen Chang\end{center}

\noindent\textbf{Abstract:} In preference logic, preference is modeled as a binary relation between possible worlds. When we discuss preference relations between propositions, we also need to consider how to lift this relation between possible worlds to propositions. Here, propositions are usually represented by their extensions, that is, sets of possible worlds. There are several choices of lifting scheme: for example, one may require even the worst world in one set to be better than the best world in another, or merely require that every world in one set have some better world in the other. The framing effect in behavioral economics shows that the lifting scheme can change with the way a question is posed or with psychological factors. This paper proves that such changes cannot be reduced to changes in preferences between possible worlds. Accordingly, we treat the lifting scheme as an independent dimension of evaluation and use a hybrid language to describe switches between lifting schemes. We thereby present a language for preference logic with dynamic lifting and prove its soundness and completeness.

\noindent\textbf{Keywords:} dynamic preference logic; preference lifting; framing effect; \mbox{completeness}

\section{Introduction}

\subsection{Overview}

In preference logic, a common approach is to represent preference as a binary relation between possible worlds and to obtain preference relations between propositions through a modal language. Here, a possible world is a complete outcome state relevant to the problem at hand. We use \(u \leq v\) to mean that "world \(v\) is at least as good as world \(u\)", and restrict the preference relation \(\leq\) to a preorder, that is, a reflexive and transitive relation. Since the extension of a proposition can contain multiple possible worlds, this raises the question of how to lift a binary relation between possible worlds to a binary relation between sets of worlds. Hansson's and Liu's studies of preference logic systematically discuss the connection between the evaluation of worlds and the evaluation of propositions~\cite{ref3,ref4}.

Consider the following example. A disease is spreading and affects 600 people. There are two treatment programs: the first ensures that 200 people survive; the second has a one-third probability of saving all 600 people and a two-thirds probability of saving no one. This example is based on the Asian disease problem. Tversky and Kahneman's experiment demonstrates preference reversal through different descriptions in terms of survival and death~\cite{ref1}. When the problem is described as "the first program will result in 400 deaths, while the second may result in no deaths or in everyone dying", people tend to choose the second program. When it is instead described as "the first program will result in 200 people surviving, while the second may result in no one surviving or in everyone surviving", people tend to choose the first program. Li and Xie further analyze this phenomenon in terms of comparisons of the best and worst possible outcomes~\cite{ref2}. In behavioral economics, this phenomenon is usually explained by the framing effect. In this example, people's preference relation between possible worlds has not changed: the world in which everyone survives is better than the world in which 200 people survive, which is in turn better than the world in which no one survives. What changes, however, is the scheme for lifting preferences from possible worlds to propositions. Under a description emphasizing death, people are more inclined to seek for opportunities, comparing the worlds with the best outcomes. While under a description emphasizing survival, they are more inclined to be risk-averse that compare the worlds with the worst outcomes. 

This example shows that the lifting schemes of people's preferences could change by the influence of some events. Accordingly, this paper focuses on dynamic changes in lifting schemes. We treat the lifting scheme as a second semantic dimension and present a preference logic in which lifting schemes change dynamically. We give its syntax and semantics and prove its soundness and completeness. In addition, we prove that the dynamics of lifting schemes cannot be reduced to the dynamics of the underlying preference relation, and examine questions such as the preservation of reflexivity and transitivity by lifting schemes.

The paper is organized as follows. Section 2 gives the syntax and semantics of preference logic with dynamic lifting. Section 3 investigates the irreducibility of lifting, first-order representation, and preservation of order properties. Section 4 establishes the axiom system and proves soundness and completeness.

\subsection{Related Work}

Preference logic provides direct quantified expressions for comparing sets. In~\cite{ref4}, propositional preferences are defined through different combinations of quantifiers, including the AA, AE, EA, and EE schemes and their modal formulations. For the AE combination, the Hoare-type lifting requires each object on the left to have a witness on the right that is at least as good, whereas the Smyth-type lifting imposes the corresponding requirement from the other side. The Egli--Milner comparison is the conjunction of these two requirements~\cite{ref17,ref18}. Comparisons on differences between extensions are further analyzed in~\cite{ref19}, where transitivity is maintained through transitive closure. The logic of convex order and its connections with other liftings are studied in~\cite{ref20}. Halpern also discusses strict comparisons on infinite sets~\cite{ref24}.

Research on ranking sets primarily examines which outcomes enter into evaluation and how these outcomes determine which set is better. Extreme-outcome rules are based on the best and worst outcomes, or assign different priorities to the two extremes. Axiomatic characterizations of Minmax, Maxmin, and related lexicographic rules are given in~\cite{ref10,ref11}. Other lexicographic comparisons of outcomes are provided in~\cite{ref12,ref39}, while~\cite{ref13} bases comparisons on median objects. Set extensions in social choice also directly specify comparisons of objects across sets: Kelly's extension requires the comparison to hold for every pair of objects across the two sets; Fishburn-type comparisons distinguish the parts exclusive to each set from their common part; and G\"ardenfors-type comparisons distinguish cases according to set inclusion~\cite{ref21,ref22,ref23}. These rules show that the treatment of the common part and the requirements on comparisons between outcomes also determine the structure of set preferences.

Injective matching and quantitative aggregation can also be used to express set lifting. Relation-preserving injective matching is used in~\cite{ref26}. Preference lifting is treated through counting-based matching in~\cite{ref27,ref28}, while~\cite{ref29} discusses the role of background conditions in lifting. Such preference lifting schemes are generally not first-order definable. In more general structural studies, operations for combining preference relations are examined in~\cite{ref30}. Work on set lifting and modal semantics in~\cite{ref31} further places these questions within a unified perspective on relational patterns and set semantics, while~\cite{ref32} investigates conditions for representing relations or distances through predicate liftings.

These studies reveal the diversity of preference lifting in terms of axiomatic constraints, specific rules, and general representation structures. Table 1.1 summarizes representative preference lifting schemes from this literature.

\clearpage
\begingroup
\footnotesize
\setlength{\tabcolsep}{4pt}
\renewcommand{\arraystretch}{1.18}
\begin{longtable}{@{}p{0.33\textwidth}p{0.62\textwidth}@{}}
\toprule\noalign{}
\raggedright Preference lifting scheme & \raggedright\arraybackslash Formal definition \\
\midrule\noalign{}
\endhead
\bottomrule\noalign{}
\endlastfoot
AA (Kelly type)~\cite{ref4,ref21} & \(\forall a \in A\mspace{6mu}\forall b \in B\mspace{6mu}(a \leq b)\) \\
AE (Hoare type)~\cite{ref4,ref19} & \(\forall a \in A\mspace{6mu}\exists b \in B\mspace{6mu}(a \leq b)\) \\
AE* (Smyth type)~\cite{ref4,ref19} & \(\forall b \in B\mspace{6mu}\exists a \in A\mspace{6mu}(a \leq b)\) \\
EA ~\cite{ref4} & \(\exists b \in B\mspace{6mu}\forall a \in A\mspace{6mu}(a \leq b)\) \\
EA* ~\cite{ref4} & \(\exists a \in A\mspace{6mu}\forall b \in B\mspace{6mu}(a \leq b)\) \\
EE ~\cite{ref4} & \(\exists a \in A\mspace{6mu}\exists b \in B\mspace{6mu}(a \leq b)\) \\
Egli--Milner ~\cite{ref17,ref18,ref19,ref20} & \(AE(A,B) \land {AE}^{*}(A,B)\) \\
Greatest-element comparison~\cite{ref6} & \(\exists a \in A (\forall x \in A ( x \leq a )) \wedge (\forall b \in B(\forall y \in B (y \leq b)) \rightarrow a \leq b)\) \\
Least-element comparison~\cite{ref6} & \(\exists a \in A (\forall x \in A ( a \leq x )) \wedge (\forall b \in B(\forall y \in B (b \leq y)) \rightarrow a \leq b)\) \\
Minmax~\cite{ref10,ref11} & \((Max(A)<Max(B))\vee(Max(A)\sim Max(B) \wedge Min(A) \leq Min(B))\)\\
Maxmin~\cite{ref10,ref11} & \((Min(A)<Min(B))\vee(Min(A)\sim Min(B) \wedge Max(A) \leq Max(B))\)\\
Leximax (Maly)~\cite{ref39} & \((A=B) \vee \exists x( x\in B\setminus A \wedge \forall y(
x<y \rightarrow (y\in A\leftrightarrow y\in B)))\)\\
Brewka type I (AE) ~\cite{ref19} & \(AE(A \setminus B,B\setminus A)\) \\
Brewka type II (AE) ~\cite{ref19} & \((A = B)\vee (A \setminus B \neq \emptyset) \wedge AE(A \setminus B, B\setminus A)\) \\
Fishburn ~\cite{ref21} & \(\forall x\forall y((x\in A\wedge x\notin B\wedge y\in B)\rightarrow x\leq y)\wedge \forall x\forall y((x\in A\wedge y\in B\wedge y\notin A)\rightarrow x\leq y)\)\\
G\"ardenfors ~\cite{ref21} & \(\bigl(A\subset B \wedge \forall x((x\in B \wedge x\notin A)\rightarrow
\forall y(y\in A\rightarrow y\leq x))\bigr) \vee
\bigl(B\subset A \wedge \forall x((x\in A \wedge x\notin B)\rightarrow\forall y(y\in B\rightarrow x\leq y))\bigr) \vee
\bigl(\neg(A\subset B) \wedge \neg(B\subset A) \wedge \forall x\forall y((x\in A \wedge x\notin B \wedge y\in B \wedge y\notin A) \rightarrow x\leq y)\bigr)\)\\
Injective matching~\cite{ref26} & \(\forall a \in A\left( a \leq f(a) \right)\), where $f$ is an injection $A \rightarrow B $\\
Counting type ~\cite{ref27,ref28} & \(\left| \{(a,b) \in A \times B:a \leq b\} \right| \geq k\)\\
\end{longtable}
\endgroup
\begin{center}\small Table 1.1: Common Preference Lifting Schemes\end{center}

\section{Syntax and Semantics}

In this section, we define the syntax and semantics of the preference logic with dynamic lifting schemes. We formulate this logic using a two-dimensional framework, with lifting schemes as the second dimension. The truth of a formula depends on the model, the world, and the lifting scheme, as expressed by \(M,w,l\vDash \varphi\); formulas are evaluated at two-dimensional points \((w,l)\).

A two-dimensional frame for preference lifting logic has two layers of structure. The first is the preference relation: $W$ is a nonempty set of possible worlds, and the binary relation $R$ on $W$ expresses that one possible world is at least as good as another. The second is $Lf$, the collection of lifting schemes from possible worlds to propositions.

Thus, a two-dimensional preference frame is denoted by
\[F = \left\langle W,Lf,R \right\rangle.\]
where \(W \neq \varnothing\), \(Lf \neq \varnothing\), and \(\ R \subseteq W \times W\).

\subsection{Syntax}

We use the static language \(L_{p}\) to reason about the frames above. \(L_{p}\) contains formulas and relation terms. Define the following pairwise disjoint sets:

\begin{center}
The set of propositional atoms \(Prop = \{ p,q,...\},\)

The set of world nominals \(Wnom = \{ i,j,...\},\)

The set of world variables \(Wvar = \{ x,y,...\},\)

The set of lifting-scheme nominals \(Lnom = \{ k,m,...\},\)

The set of lifting-scheme variables \(\ Lvar\  = \ \{\lambda,\mu,...\}.\)
\end{center}

The operators \(@_{(s,t)},\lbrack\rho\rbrack, \downarrow (x,\lambda)\) are modal operators of \(L_{p}\), with the following meanings:

\begin{center}
\(@_{(s,t)}\varphi\): \(\varphi\) holds after moving the current evaluation point to \((s,t)\).

\(\downarrow (x,\lambda)\varphi\): \(\varphi\) holds after assigning the current evaluation point to \((x,\lambda)\).

\(\lbrack\rho\rbrack\varphi\): \(\varphi\) holds at every world accessible from the current world via the relation \(\rho\).
\end{center}

Here, \(s \in Wnom \cup Wvar\), \(t \in Lnom \cup Lvar,x \in Wvar,\lambda \in Lvar\).

Formulas \(\varphi\) and relation terms \(\rho\) are then defined by simultaneous recursion as follows:\footnote{We use the relation construction $?\varphi$ in order to express the reduction axioms for further studies. As ~\cite{ref5} has shown, the dynamic operators of prefernece update could be axiomize if it's non-iteration PDL definable.}

\[\varphi: = p \mid i \mid k \mid x \mid \lambda \mid \neg\varphi \mid (\varphi \land \psi) \mid @_{(s,t)}\varphi \mid \  \downarrow (x,\lambda)\varphi \mid \lbrack\rho\rbrack\varphi,\]
\[\rho: = r \mid - \rho \mid \rho^{- 1} \mid \rho \cup \sigma \mid \rho;\sigma \mid \ ?\varphi.\]

\textbf{Definition 2.1:} Let \(Fresh:L \rightarrow P(Wvar \cup Lvar)\) denote the set of variables that do not occur in a formula or relation term. We say that a variable \(a \in Fresh(\varphi)\) if and only if \(a\) does not occur in the string constituting \(\varphi\). Similarly, \(a \in Fresh(\rho)\) if and only if \(a\) does not occur in the string constituting \(\rho\).

Let \(Fv:L \rightarrow P(Wvar \cup Lvar)\) denote the set of variables that \textbf{occur free} in a formula or relation term. We say that \(a\) \textbf{occurs free} in \(\varphi\) if and only if \(a \in Fv(\varphi)\); similarly, \(a\) \textbf{occurs free} in \(\rho\) if and only if \(a \in Fv(\rho)\).

\(Fv(\varphi)\) and \(Fv(\rho)\) are defined by simultaneous recursion as follows:\\
Formulas:
\[
\begin{aligned}
&Fv(p) = Fv(i) = Fv(k): = \varnothing, \\
&Fv(x): = \{ x\}, \\
&Fv(\lambda): = \{\lambda\}, \\
&Fv(\neg\varphi): = Fv(\varphi), \\
&Fv(\varphi \land \psi): = Fv(\psi) \cup Fv(\varphi), \\
&Fv(@_{(s,t)}\varphi): = Fv(\varphi) \cup (\{ s,t\} \cap (Wvar \cup Lvar)), \\
&Fv( \downarrow (x,\lambda)\varphi): = Fv(\varphi)\backslash\{ x,\lambda\}, \\
&Fv(\lbrack\rho\rbrack\varphi): = Fv(\rho) \cup Fv(\varphi);
\end{aligned}
\]
Relation terms:
\[
\begin{aligned}
&Fv(r): = \varnothing, \\
&Fv( - \rho): = Fv(\rho), \\
&Fv(\rho^{- 1}): = Fv(\rho), \\
&Fv(\rho \cup \sigma): = Fv(\rho) \cup Fv(\sigma), \\
&Fv(\rho;\sigma): = Fv(\rho) \cup Fv(\sigma), \\
&Fv(?\varphi): = Fv(\varphi).
\end{aligned}
\]

\textbf{Definition 2.2:} Let the simultaneous recursion depth \(rk:L_{p} \rightarrow N\) be defined recursively as follows:\\
Formulas:
\[
\begin{aligned}
&rk(p) = rk(i) = rk(k) = rk(x) = rk(\lambda): = 0, \\
&rk(\neg\varphi): = rk(\varphi) + 1, \\
&rk(\varphi \land \psi): = max(rk(\psi),rk(\varphi)) + 1, \\
&rk(@_{(s,t)}\varphi): = rk(\varphi) + 1, \\
&rk( \downarrow (x,\lambda)\varphi): = rk(\varphi) + 1, \\
&rk(\lbrack\rho\rbrack\varphi): = max(rk(\rho),rk(\varphi)) + 1;
\end{aligned}
\]
Relation terms:
\[
\begin{aligned}
&rk(r): = 0, \\
&rk( - \rho): = rk(\rho) + 1, \\
&rk(\rho^{- 1}): = rk(\rho) + 1, \\
&rk(\rho \cup \sigma): = max(rk(\rho),rk(\sigma)) + 1, \\
&rk(\rho;\sigma): = max(rk(\rho),rk(\sigma)) + 1, \\
&rk(?\varphi): = rk(\varphi) + 1.
\end{aligned}
\]

\textbf{Proposition 2.3:} The definition of \(L_{p}\) is well-founded.

\textbf{Proof:} By Definition 2.2, the mutually dependent formation rules satisfy \(rk(?\varphi) > rk(\varphi),rk(\lbrack\rho\rbrack\varphi) > rk(\rho)\). Hence the definition is well-founded. \hfill\(\square\)

\textbf{Definition 2.4:} Substitution for free variables is denoted by

\[
\begin{aligned}
&\varphi\lbrack s/x,t/\lambda\rbrack, \\
&\rho\lbrack s/x,t/\lambda\rbrack.
\end{aligned}
\]
where \(s \in Wnom \cup Wvar\), \(t \in Lnom \cup Lvar\),
\(x \in Wvar \cap (Fresh(\varphi) \cup Fv(\varphi))\) (or \(Fresh(\rho) \cup Fv(\rho)\)),
and \(\lambda \in Lvar \cap (Fresh(\varphi) \cup Fv(\varphi))\) (or \(Fresh(\rho) \cup Fv(\rho)\)).\\
The recursive definition is as follows:

Formulas:
\[
\begin{aligned}
&p\lbrack s/x,t/\lambda\rbrack: = p, \\
&i\lbrack s/x,t/\lambda\rbrack: = i, \\
&y\lbrack s/x,t/\lambda\rbrack: = y, \\
&x\lbrack s/x,t/\lambda\rbrack: = s, \\
&k\lbrack s/x,t/\lambda\rbrack: = k, \\
&\mu\lbrack s/x,t/\lambda\rbrack: = \mu, \\
&\lambda\lbrack s/x,t/\lambda\rbrack: = t, \\
&(\neg\varphi)\lbrack s/x,t/\lambda\rbrack: = \neg(\varphi\lbrack s/x,t/\lambda\rbrack), \\
&(\varphi \land \psi)\lbrack s/x,t/\lambda\rbrack: = \varphi\lbrack s/x,t/\lambda\rbrack \land \psi\lbrack s/x,t/\lambda\rbrack, \\
&(@_{(u,v)}\varphi)\lbrack s/x,t/\lambda\rbrack: = @_{(u\lbrack s/x\rbrack,v\lbrack t/\lambda\rbrack)}\varphi\lbrack s/x,t/\lambda\rbrack, \\
&( \downarrow (y,\mu)\varphi)\lbrack s/x,t/\lambda\rbrack: = \downarrow (y',\mu')((\varphi\lbrack y'/y,\mu'/\mu\rbrack)\lbrack s/x,t/\lambda\rbrack), \\
&(\lbrack\rho\rbrack\varphi)\lbrack s/x,t/\lambda\rbrack: = \lbrack\rho\lbrack s/x,t/\lambda\rbrack\rbrack(\varphi\lbrack s/x,t/\lambda\rbrack).
\end{aligned}
\]
Here, \((y',\mu')\) satisfy \(y' \in Wvar \cap Fresh(\varphi)\),
\(\mu' \in Lvar \cap Fresh(\varphi)\), and
\(y' \neq x\), \( y' \neq s\), \(\mu' \neq \lambda\), \(\mu' \neq t\).

Relation terms:
\[
\begin{aligned}
&r\lbrack s/x,t/\lambda\rbrack: = r, \\
&( - \rho)\lbrack s/x,t/\lambda\rbrack: = - (\rho\lbrack s/x,t/\lambda\rbrack), \\
&(\rho^{- 1})\lbrack s/x,t/\lambda\rbrack: = {(\rho\lbrack s/x,t/\lambda\rbrack)}^{- 1}, \\
&(\rho \cup \sigma)\lbrack s/x,t/\lambda\rbrack: = \rho\lbrack s/x,t/\lambda\rbrack \cup \sigma\lbrack s/x,t/\lambda\rbrack, \\
&(\rho;\sigma)\lbrack s/x,t/\lambda\rbrack: = \rho\lbrack s/x,t/\lambda\rbrack;\sigma\lbrack s/x,t/\lambda\rbrack, \\
&(?\varphi)\lbrack s/x,t/\lambda\rbrack: = ?(\varphi\lbrack s/x,t/\lambda\rbrack).
\end{aligned}
\]

\subsection{Semantics}

Let \(V_{p}:Prop \rightarrow P(W)\), \(\eta_{W}:Wnom \rightarrow W\), and
\(\eta_{L}:Lnom \rightarrow L\) interpret the propositional atoms and nominals.
A model for the preference logic with dynamic lifting is then \(M= \left \langle F,V \right \rangle\), where
\(V := \left\langle V_{p},\eta_{W},\eta_{L} \right\rangle\).
Let \(g_{W}:Wvar \rightarrow W\) and \(g_{L}:Lvar \rightarrow L\) be assignments to the variables, and let
\(g: = \left\langle g_{W},g_{L} \right\rangle\). We can now give the semantic interpretation of the preference logic with dynamic lifting as follows.

\textbf{Definition 2.5:} Given a model \(M = \left\langle F,V \right\rangle\), the satisfaction relation \(M,g,w,l \vDash \varphi\) for formulas \(\varphi\) and the interpretation \(\llbracket \rho \rrbracket^{M,g,l} \in P(W \times W)\) of relation terms \(\rho\) are defined by simultaneous recursion as follows.\\
Formulas:
\[
\begin{aligned}
&M,g,w,l \vDash p \Longleftrightarrow w \in V(p), \\
&M,g,w,l \vDash i \Longleftrightarrow (w = \eta_{W}(i)), \\
&M,g,w,l \vDash x \Longleftrightarrow (w = g_{W}(x)), \\
&M,g,w,l \vDash k \Longleftrightarrow (l = \eta_{L}(k)), \\
&M,g,w,l \vDash \lambda \Longleftrightarrow (l = g_{L}(\lambda)), \\
&M,g,\ w,\ l \vDash \neg\varphi \Longleftrightarrow M,g,\ w,\ l \nvDash \varphi, \\
&M,g,w,l \vDash (\varphi \land \psi) \Longleftrightarrow M,g,w,l \vDash \varphi\ \&\ M,g,w,l \vDash \psi, \\
&M,g,\ w,\ l \vDash @_{(s,t)}\varphi \Longleftrightarrow M,g,\llbracket s \rrbracket^{M,g},\llbracket t \rrbracket^{M,g} \vDash \varphi, \\
&M,g,w,l \vDash \downarrow (x,\lambda)\varphi \Longleftrightarrow M,g\lbrack x \mapsto w,\lambda \mapsto l\rbrack,w,l \vDash \varphi, \\
&M,g,w,l \vDash \lbrack\rho\rbrack\varphi \Longleftrightarrow \forall v \in W\left( (w,v) \in \llbracket \rho \rrbracket^{M,g,l} \Rightarrow M,g,v,l \vDash \varphi \right).
\end{aligned}
\]

Here, \(\llbracket \  \rrbracket^{M,g}:Wnom \cup Lnom \cup Wvar \cup Lvar \rightarrow W \cup L\) interprets the nominals and variables, with \(\llbracket i \rrbracket^{M,g}: = \eta_{W}(i),\llbracket x \rrbracket^{M,g}: = g_{W}(x),\llbracket k \rrbracket^{M,g}: = \eta_{L}(k),\llbracket \lambda \rrbracket^{M,g}: = g_{L}(\lambda);\)\\
Relation terms:
\[
\begin{aligned}
&\llbracket r \rrbracket^{M,g,l}: = R, \\
&\llbracket  - \rho \rrbracket^{M,g,l}: = \left\{ (u,v) \in W \times W \middle| (u,v) \notin \llbracket \rho \rrbracket^{M,g,l} \right\}, \\
&\llbracket \rho^{- 1} \rrbracket^{M,g,l}: = \left\{ (u,v) \in W \times W \middle| (v,u) \in \llbracket \rho \rrbracket^{M,g,l} \right\}, \\
&\llbracket \rho \cup \sigma \rrbracket^{M,g,l}: = \llbracket \rho \rrbracket^{M,g,l} \cup \llbracket \sigma \rrbracket^{M,g,l}, \\
&\llbracket \rho;\sigma \rrbracket^{M,g,l}: = \left\{ (u,v) \in W \times W \middle| \exists z \in W,(u,z) \in \llbracket \rho \rrbracket^{M,g,l} \land (z,v) \in \llbracket \sigma \rrbracket^{M,g,l} \right\}, \\
&\llbracket ?\varphi \rrbracket^{M,g,l}: = \left\{ (u,u) \in W \times W \middle| M,g,u,l \vDash \varphi \right\}.
\end{aligned}
\]

Proposition 2.6 and Lemma 2.7 below can be proved by induction on the simultaneous recursion depth, since every constituent in the recursive definition has strictly smaller recursion depth than the expression it constitutes.

\textbf{Proposition 2.6 (Well-definedness of the semantics):} For every model \(M\), assignment \(g\), evaluation point \(w,l\), formula \(\varphi\), and relation term \(\rho\), either \(M,g,w,l \vDash \varphi\) or \(M,g,w,l \nvDash \varphi\), and \(\llbracket \rho \rrbracket^{M,g,l} \in P(W \times W)\).

\textbf{Lemma 2.7 (Substitution lemma for free variables):} For any \(s \in Wnom \cup Wvar\), \(t \in Lnom \cup Lvar\), \(x \in Wvar\), and \(\lambda \in Lvar\), and for every model \(M\), assignment \(g\), evaluation point \(w,l\), formula \(\varphi\), and relation term \(\rho\), we have:
\[
\begin{aligned}
&M,g,w,l \vDash \varphi\lbrack s/x,t/\lambda\rbrack \Longleftrightarrow M,g\left\lbrack x \mapsto \llbracket s \rrbracket^{M,g},\lambda \mapsto \llbracket t \rrbracket^{M,g} \right\rbrack,w,l \vDash \varphi, \\
&\llbracket \rho\lbrack s/x,t/\lambda\rbrack \rrbracket^{M,g,l} = \llbracket \rho \rrbracket^{M,g\left\lbrack x \mapsto \llbracket s \rrbracket^{M,g},\lambda \mapsto \llbracket t \rrbracket^{M,g} \right\rbrack,l}.
\end{aligned}
\]

\section{Preference Lifting}

In this section, we define preference lifting and preference between propositions. We show that the dynamics of preference lifting schemes cannot be reduced to preference dynamics. We also show that all preferences between propositions obtained from first-order definable preference lifting schemes can be regarded as definitional extensions of \(L_{p}\). We conclude this section by discussing some common preference lifting schemes.

\subsection{Preference Lifting and Its Representation}

A preference lifting scheme \(l\) is a function \(P(W \times W) \rightarrow P(P(W)\times P(W))\). We write \(A\, l(R)\, B\) for \((A,B) \in l_{W}(R)\). We call a preference lifting scheme singleton-conservative if and only if, for any \(a,b \in W\), we have \(\{ a\}\, l(R)\,\{ b\} \Leftrightarrow aRb\). Singleton conservativity is a natural requirement on preference lifting.

The truth value of a preference relation \(\varphi \leq \psi\) between propositions depends only on the extensions of \(\varphi , \psi\), the preference relation \(leq\), and the preference lifting scheme \(l\). That is,
\[M,g,w,l \vDash \varphi \leq \psi \Leftrightarrow \lbrack\varphi\rbrack^{M,g,l}\, l(R)\,\lbrack\psi\rbrack^{M,g,l}.\]

For singleton-conservative preference liftings, the dynamics of lifting schemes cannot be reduced to the dynamics of the preference relation between possible worlds itself. We have the following theorem.

\textbf{Theorem 3.1 (Irreducibility):} Let \(l,l'\) be two singleton-conservative preference lifting schemes. If, on the same domain of worlds, \(Al(R')B \Leftrightarrow Al'(R)B\) holds for all \(A,B,R\), then \(R' = R\) and \(l = l'\).

\textbf{Proof:} For all singleton sets \(\{a\}\) and \(\{ b\}\), singleton conservativity gives \(\{a\}l(R')\{ b\}\) if and only if \(aR'b\), and \(\{a\}l'(R)\{ b\}\) if and only if \(aRb\). Since \(\{a\}l(R')\{b\} \Leftrightarrow \{a\}l'(R)\{b\}\), we have \(aR'b \Leftrightarrow aRb\) for any \(a,b \in W\). Hence \(R=R'\). The premise is then equivalent to \(Al(R)B \Leftrightarrow Al'(R)B\) for all \(A,B,R\), so \(l = l'\) by definition. \hfill\(\square\)

We study first-order definable preference liftings.

\textbf{Definition 3.2:} We say that a preference lifting scheme \(l\) is first-order definable if and only if there is a first-order formula with no free variables, generated by
\[\theta:= x \leq y \mid x = y \mid x \in A \mid x \in B \mid \neg\theta \mid (\theta \land \theta) \mid \exists x\,\theta.\]
such that, for every nonempty set of worlds \(W\), nonempty sets \(A,B \subseteq W\), and binary relation \(R\) on \(W\),
\[A\, l(R)\, B \Leftrightarrow (W,R,A,B) \vDash \theta.\]
We call \(\theta\) a first-order expression for the preference lifting scheme \(l\).

Every first-order definable preference lifting can be expressed in \(L_{p}\). This can be proved by constructing a translation function. We have the following theorem.

\textbf{Theorem 3.3:} For any first-order definable lifting \(l\), let \(\theta\) be its first-order expression. Then there is a formula \(\delta(\varphi,\psi)\) in \(L_{p}\) such that
\[M,g,w,t \vDash \delta(\varphi,\psi) \Leftrightarrow \lbrack\varphi\rbrack^{M,g,t}\, l(R)\,\lbrack\psi\rbrack^{M,g,t}.\]

\textbf{Proof:} This amounts to proving that \(M,g,w,t \vDash \delta(\varphi,\psi) \Leftrightarrow W, R, \lbrack\varphi\rbrack^{M,g,t}, \lbrack\psi\rbrack^{M,g,t} \vDash \theta\). Let \(x,y,z\) be free world variables that do not occur in \(\varphi,\psi\), and let \(\lambda\) be a free world variable that does not occur in \(\varphi,\psi\). We can construct a translation function \(T_{\varphi,\psi}^{\lambda}\) from first-order expressions into \(L_{p}\) as follows.
\[
\begin{gathered}
T_{\varphi,\psi}^{\lambda}(x \in A) = @_{(x,\lambda)}\varphi,\\
T_{\varphi,\psi}^{\lambda}(x \in B) = @_{(x,\lambda)}\psi,\\
T_{\varphi,\psi}^{\lambda}(x = y) = @_{(x,\lambda)}y,\\
T_{\varphi,\psi}^{\lambda}(x \leq y) = @_{(x,\lambda)}\langle r\rangle y,\\
T_{\varphi,\psi}^{\lambda}(\neg\alpha) = \neg T_{\varphi,\psi}^{\lambda}(\alpha),\\
T_{\varphi,\psi}^{\lambda}(\alpha \land \beta) = T_{\varphi,\psi}^{\lambda}(\alpha) \land T_{\varphi,\psi}^{\lambda}(\beta),\\
T_{\varphi,\psi}^{\lambda}(\exists x\,\alpha) = \neg\lbrack r \cup - r\rbrack\neg\left( \downarrow (x,\mu)T_{\varphi,\psi}^{\lambda}(\alpha) \right).
\end{gathered}
\]
Here, \(\mu\) is a lifting-scheme variable distinct from \(\lambda\) that does not occur in \(\alpha\). Thus, the expression for the lifting scheme \(l\) in \(L_{p}\) is
\[\delta(\varphi,\psi): = \downarrow (z,\lambda)T_{\varphi,\psi}^{\lambda}\left( \theta \right).\]
\hfill\(\square\)

Let \(Lf = \{ l_{0},l_{1},\ldots\}\), and let \(k_{i}\) denote \(l_{i}\). We may then set
\[\varphi \leq \psi := \bigvee_{i < \omega}(k_{i} \wedge \delta_{i}(\varphi,\psi)).\]

\subsection{Preservation of Preorders by Preference Liftings}

If preference between possible worlds is understood as ``world $u$ is at least as good as world $v$,'' this relation should be a preorder, that is, a reflexive and transitive relation. We would like the preference relation between propositions to be a preorder as well: a proposition $\varphi$ is at least as good as itself; and if $\varphi$ is at least as good as $\psi$, and $\psi$ is at least as good as $\theta$, then $\varphi$ is at least as good as $\theta$. We therefore need to examine whether preference liftings preserve preorders.

\textbf{Definition 3.4 (Preservation of Order Properties):} We call \(l\) reflexivity-preserving if and only if, for any $W$, $R \in P(W \times W)$, if \(\forall a \in W, aRa\), then \(\forall A \subseteq W, A \neq \emptyset \rightarrow Al(r)A \). We call \(l\) transitivity-preserving if and only if, for any $W$, $R \in P(W \times W)$, if \(\forall a,b,c \in W, (aRb \wedge bRc) \rightarrow aRc\), then \(\forall A,B,C \subseteq W, A,B,C \neq \emptyset \rightarrow ( Al(r)B \wedge Bl(r)C \rightarrow Al(r)C )\). We call \(l\) preorder-preserving if and only if \(l\) is both reflexivity-preserving and transitivity-preserving.

We write \(Pre \vDash \alpha\) to mean that \((W,R) \vDash \alpha\) for every nonempty $W$ and every preorder relation $R$ on $W$. Lavrov's results on the relevant elementary theories show that the first-order validity problem for preorders is undecidable~\cite{ref38}. That is, there is no uniform algorithm that, given an arbitrary first-order sentence $\alpha$, decides whether \(Pre \vDash \alpha\). On the basis of this result, we have the following theorem.

\textbf{Theorem 3.5 (Undecidability of Order-Property Preservation):} There is no algorithm that decides, for an arbitrary preference lifting scheme $l$, whether it is reflexivity-preserving. Likewise, there is no algorithm that decides, for an arbitrary preference lifting scheme $l$, whether it is transitivity-preserving.

\textbf{Proof:}
First, we show that there is no uniform decision procedure for preservation of reflexivity. For any sentence $\alpha$, we can construct a lifting scheme $l_{\alpha}$ such that $Pre\vDash \alpha \Leftrightarrow$ ($l_{\alpha }$ is reflexive). We show that we may set $l_{\alpha }(A,B)=\alpha \wedge (\exists x,x\in A)\wedge (\exists y,y\in B)$. If \(Pre \vDash \alpha\), take any preorder $(W,R)$ and any nonempty set $A$. Since $\alpha$ holds on this preorder and $(\exists x, x \in A)$ holds, $l_{\alpha }(A,A)$ holds. Conversely, suppose that \(Pre \nvDash \alpha\). Then there is a preorder $(W,R) \nvDash \alpha$, so $l_{\alpha }(A,A)$ is false on $(W,R)$. Thus, if there were a decision algorithm for preservation of reflexivity, then \(Pre \vDash \alpha\) would be decidable, a contradiction.

Next, we show that there is no uniform decision procedure for preservation of transitivity. For any sentence $\alpha$, we can construct a lifting scheme $l_{\alpha}$ such that $Pre\vDash \alpha \Leftrightarrow$ ($l_{\alpha }$ is transitive). If $\alpha$ is true on the one-world preorder, we may set $l_{\alpha }(A,B)=\alpha \vee \forall x\left(x\in A\rightarrow x\in B\right)\vee \forall y\left(y\in B\rightarrow y\in A\right)$. If \(Pre \vDash \alpha\), take any preorder $(W,R)$; then $(W,R) \vDash \alpha$. In this case, $l_{\alpha }(A,B)$ holds for any sets $A,B$, so $l_{\alpha }$ is transitive. Conversely, suppose that $Pre \nvDash \alpha$. Then there is a preorder $(W,R)\nvDash \alpha$. In this case, $W$ has at least two elements, and we may set $A=\{u\},B=\{u,v\},C=\{v\}$. We then have $l_{\alpha }(A,B)$ and $l_{\alpha }(B,C)$ but $\neg l_{\alpha }(A,C)$, so $l_{\alpha }$ is not transitive. If $\alpha$ is false on the one-world preorder, set $l_{\alpha }(A,B)=\forall x\left(x\in A\rightarrow x\in B\right)\vee \forall y\left(y\in B\rightarrow y\in A\right)$; this lifting is not transitivity-preserving.

\hfill\(\square\)

\textbf{Corollary 3.6:} There is no algorithm that decides, for an arbitrary singleton-conservative preference lifting scheme $l$, whether it is preorder-preserving.

Although there is no uniform decision procedure for whether singleton-conservative preference liftings preserve preorders, we can give a sufficient condition for preservation of preorders. For the common preference liftings listed in Table 1.1, this condition characterizes their preservation of preorders. Write
\[
\bigwedge_{i=1}^{k}\forall\overline{z_i}
\left(\eta_i(A,\overline{z_i})\rightarrow\eta_i(B,\overline{z_i})\right)\tag{CT}
\]
where, for each $i$, \(\overline{z_i}\) is a tuple of possible worlds, and \(\eta_i(X,\overline{z_i})\) is a first-order formula whose only inputs are \(X\) and \(\overline{z_i}\). We call this condition the \textbf{common-test form}. For a preference lifting scheme satisfying the common-test form, $Al(R)B$ holds if and only if, for every test $\eta_i$, whenever $A$ satisfies $\eta_i$, $B$ also satisfies that test.

\textbf{Theorem 3.7:} For any first-order definable lifting scheme \(l\), if its first-order expression can be written in $CT$ form, then \(l\) is preorder-preserving.

\textbf{Proof:} This follows from the reflexivity and transitivity of logical implication. \hfill\(\square\)

Although $CT$ is only a sufficient condition for preservation of preorders\footnote{For example, \((AE \wedge AE^{*}) \vee EA\) is preorder-preserving but cannot be characterized by $CT$.}, for common preference liftings (such as those in Table 1.1), this condition suffices to characterize their preservation of preorders. We have the following proposition.

\textbf{Proposition 3.8:} For the common preference lifting schemes listed in Table 1.1, if a scheme is first-order definable, then it is preorder-preserving if and only if its first-order expression can be written in $CT$ form.

\textbf{Proof:} Sufficiency follows from Theorem 3.7. We now prove necessity by examining Table 1.1 item by item and giving a first-order expression in $CT$ form for every preorder-preserving preference lifting.\\
AA (Kelly): Does not preserve reflexivity.\\
AE (Hoare):
\[\eta_{H}(X,z): = \exists x(x \in X \wedge z \leq x)\]
\[\mathit{A}\mathit{E}(A,B) \leftrightarrow \forall z (\eta_{H}(A,z) \rightarrow \eta_{H}(B,z))\]
AE* (Smyth):
\[\eta_{S}(X,z): = \neg\exists x(x \in X \wedge x \leq z)\]
\[{\mathit{A}\mathit{E}}^{*}(A,B) \leftrightarrow \forall z(\eta_{S}(A,z) \rightarrow \eta_{S}(B,z))\]
Egli--Milner:
\[\eta_{E}(X,z): =\eta_{H}(X,z) \wedge \eta_{S}(X,z)\]
\[A \leq_{EM} B \leftrightarrow \forall z (\eta_{E}(A,z) \rightarrow \eta_{E}(B,z))\]
EA: Does not preserve reflexivity.\\
EA*: Does not preserve reflexivity.\\
EE: Does not preserve transitivity.\\
Max: When greatest elements exist, we have
\[\eta_{Mx}(X,z): = \exists x(x \in X \wedge \forall y (y \in X \rightarrow y\leq x)) \rightarrow z \leq x\]
\[A \leq_{Mx} B \leftrightarrow \forall z (\eta_{Mx}(A,z) \rightarrow \eta_{Mx}(B,z))\]
Min: When least elements exist, we have
\[\eta_{Mn}(X,z): = \exists x(x \in X \wedge \forall y (y \in X \rightarrow x\leq y)) \rightarrow x \leq z\]
\[A \leq_{Mn} B \leftrightarrow \forall z (\eta_{Mn}(A,z) \rightarrow \eta_{Mn}(B,z))\]
Maxmin: When both greatest and least elements exist, we have
\[
\begin{aligned}
\eta_{Mxmn}(X,u,v)&: = \exists x(x \in X \wedge (u < x \vee (u \sim x \wedge \forall y (y \in X \rightarrow (y \leq u)) \\ &\qquad \wedge \exists z(z \in X \wedge (v \leq z))
\end{aligned}
\]
\[A \leq_{Mxmn} B \leftrightarrow \forall u \forall v (\eta_{Mxmn}(A,u,v) \rightarrow \eta_{Mxmn}(B,u,v))\]
Minmax: When both greatest and least elements exist, we have
\[
\begin{aligned}
\eta_{Mnmx}(X,u,v)&: = \exists x(x \in X \wedge (x < u \vee (x \sim u \wedge \forall y (y \in X \rightarrow (u \leq y)) \\ &\qquad \wedge \exists z(z \in X \wedge (z \leq v))
\end{aligned}
\]
\[A \leq_{Mnmx} B \leftrightarrow \forall u \forall v (\eta_{Mnmx}(A,u,v) \rightarrow \eta_{Mnmx}(B,u,v))\]
Leximax (Maly): Does not preserve transitivity on general preorders; it preserves preorders only on linear orders. Take $W=\{a,b,c\}$ and $R=\Delta_{W}$. Let \(A=\{a,b\}, B=\{c\}, C=\{a\}\). Then $A \leq B$ and $B \leq C$, but $\neg( A \leq C)$.\\
Brewka: Does not preserve transitivity. For the first Hoare type, take \(b < a \sim c\) and \(A=\{a,b\}, B=\{c\}, C=\{a\}\). The first two comparisons hold, whereas the final comparison fails because it requires a witness from the empty set \(C\backslash A\). For the second type, take \(a \sim b\), with $c$ incomparable with both, and let \(A=\{a\},B=\{b\},C=\{a,c\}\). The same phenomenon occurs, and transitivity fails.\\
Fishburn:\footnote{The $CT$ representation of the Fishburn-type lifting here shows that the scope of $CT$ representation differs from the conditions for functor extensions in~\cite{ref32,ref33}, since the Fishburn-type lifting is not invariant under morphisms.}
\[\eta_{1}(X,z): = z \in X \vee \forall x(x \in X \rightarrow z \leq x)\]
\[\eta_{2}(X,z): = z \notin X \wedge \exists x ( x \in X \wedge \neg(x \leq z) )\]
\[A \leq_{F} B \leftrightarrow (\forall u(\eta_1(A,u)\rightarrow\eta_1(B,u))) \wedge (\forall v(\eta_2(A,v)\rightarrow\eta_2(B,v))) \]
Gärdenfors: Does not preserve transitivity. Let \(W = \{a,b,c,d\} \) and \(R=\Delta_{W} \cup \{\leq d,b\leq c\}\). Take \(A=\{a,b\},B=\{a,c\},C=\{c,d\}\). The comparisons on the set differences hold from A to B and from B to C, respectively, but the comparison from A to C additionally requires \(a\leq c\) and \(b\leq d\), and therefore fails.\\
Preference liftings based on injective matching and counting are not first-order definable.
\par\nobreak\noindent\hspace*{\fill}\(\square\)

\section{Axiom System}

In this section, we present an axiom system for dynamic preference lifting and prove its soundness and completeness.

\subsection{Axiom System and Soundness}

Let $H_{PDL}^{2}$ denote the axiom system for the language $L_{p}$, consisting of the following axioms and inference rules.

\textbf{Propositional logic:}

\begin{equation}
\varphi \rightarrow (\psi \rightarrow \varphi)
\tag{1}
\end{equation}

\begin{equation}
(\varphi \rightarrow (\psi \rightarrow \theta)) \rightarrow ((\varphi \rightarrow \psi) \rightarrow (\varphi \rightarrow \theta))
\tag{2}
\end{equation}

\begin{equation}
(\neg\varphi \rightarrow \neg\psi) \rightarrow (\psi \rightarrow \varphi)
\tag{3}
\end{equation}

\textbf{Modal K axiom:}

For every relation term $\rho$, we have

\begin{equation}
[\rho](\varphi \rightarrow \psi) \rightarrow \left( [\rho]\varphi \rightarrow [\rho]\psi \right)
\tag{4}
\end{equation}

\textbf{Axioms for $@_{(s,t)}$:}

For every $(s,t):s \in Wnom \cup Wvar,t \in Lnom \cup Lvar$, we have the following axioms.

K axiom:

\begin{equation}
@_{(s,t)}(\varphi \rightarrow \psi) \rightarrow \left( @_{(s,t)}\varphi \rightarrow @_{(s,t)}\psi \right)
\tag{5}
\end{equation}

Self-duality:

\begin{equation}
@_{(s,t)}\varphi \leftrightarrow \neg@_{(s,t)}\neg\varphi
\tag{6}
\end{equation}

Self-naming:

\begin{equation}
@_{(s,t)}(s \land t)
\tag{7}
\end{equation}

Agree:

\begin{equation}
@_{(u,v)}@_{(s,t)}\varphi \leftrightarrow @_{(s,t)}\varphi
\tag{8}
\end{equation}

Intro:

\begin{equation}
(s \land t) \rightarrow (@_{(s,t)}\varphi \leftrightarrow \varphi)
\tag{9}
\end{equation}

Back:

\begin{equation}
\left\langle \rho \right\rangle@_{(s,t)}\varphi \rightarrow @_{(s,t)}\varphi\qquad\text{(for every }\rho\text{)}
\tag{10}
\end{equation}

\textbf{Axiom for $\downarrow(x,\lambda)$:}

When $x,\lambda \in Fresh(\varphi) \cup Fv(\varphi)$, we have:

\begin{equation}
@_{(s,t)}( \downarrow (x,\lambda)\varphi \leftrightarrow \varphi[ s/x,t/\lambda])\qquad\text{(DA)}
\tag{11}
\end{equation}

\textbf{Axioms for relation terms:}

For $s,u \in Wnom \cup Wvar$ and $t \in Lnom \cup Lvar$, we have:

\begin{equation}
@_{(s,t)}\left\langle - \rho \right\rangle u \leftrightarrow \neg@_{(s,t)}\left\langle \rho \right\rangle u\qquad\text{(Comp)}
\tag{12}
\end{equation}

\begin{equation}
@_{(s,t)}\left\langle \rho^{- 1} \right\rangle u \leftrightarrow @_{(u,t)}\left\langle \rho \right\rangle s\qquad\text{(Conv)}
\tag{13}
\end{equation}

\begin{equation}
((@_{(s,t)}\left\langle \rho \right\rangle u) \land @_{(u,t)}\varphi) \rightarrow (@_{(s,t)}\left\langle \rho \right\rangle\varphi)\qquad\text{(Bridge)}
\tag{14}
\end{equation}

\begin{equation}
[\rho \cup \sigma]\varphi \leftrightarrow ([\rho]\varphi \land [\sigma]\varphi)\qquad\text{(Choice)}
\tag{15}
\end{equation}

\begin{equation}
[\rho;\sigma]\varphi \leftrightarrow [\rho][\sigma]\varphi\qquad\text{(Seq)}
\tag{16}
\end{equation}

\begin{equation}
[?\psi]\varphi \leftrightarrow (\psi \rightarrow \varphi)\qquad\text{(Test)}
\tag{17}
\end{equation}

\textbf{Rigidity axioms:}\par\nobreak

For $p \in Prop$, $s,u \in Wnom \cup Wvar$, and $t,v \in Lnom \cup Lvar$, we have:

\begin{equation}
@_{(s,t)}p \leftrightarrow @_{(s,v)}p
\tag{18}
\end{equation}

\begin{equation}
@_{(s,t)}u \leftrightarrow @_{(s,v)}u
\tag{19}
\end{equation}

\begin{equation}
@_{(s,t)}v \leftrightarrow @_{(u,t)}v
\tag{20}
\end{equation}

\begin{equation}
@_{(s,t)}\left\langle r \right\rangle u \leftrightarrow @_{(s,v)}\left\langle r \right\rangle u
\tag{21}
\end{equation}

\textbf{Inference rules:}

\[
\frac{\varphi\qquad\varphi\rightarrow\psi}{\psi}
\qquad\text{(MP)}
\]

If $\vdash\varphi$, then we have
\[
\frac{\varphi}{[\rho]\varphi}
\qquad\text{(Nec}\mathord{\Box}\text{)}
\]
and
\[
\frac{\varphi}{@_{(s,t)}\varphi}
\qquad\text{(Nec@)}
\]

If neither the world nominal $i$ nor the lifting-scheme nominal $k$ occurs in $\varphi$, then we have
\[
\frac{@_{(i,k)}\varphi}{\varphi}
\qquad\text{(Name)}
\]

If the world nominal $u$ does not occur in $\rho,\varphi,s,t$, then we have
\[
\frac{@_{(s,t)}\langle\rho\rangle u\rightarrow @_{(u,t)}\varphi}
{@_{(s,t)}[\rho]\varphi}
\qquad\text{(BG)}
\]

The following two rules substitute terms of the same sort throughout a theorem, avoiding variable capture:
\[
\frac{\chi}{\chi[x/i]}
\qquad\text{(SubW)}
\]
\[
\frac{\chi}{\chi[\lambda/k]}
\qquad\text{(SubL)}
\]
where $\vdash\chi$, $i\in Wnom$, $k\in Lnom$, $x\in Wvar$, and $\lambda\in Lvar$; $[x/i]$ and $[\lambda/k]$ denote the uniform replacement of all occurrences of the nominals $i$ and $k$, respectively, by variables of the same sort.

We have the following theorem.\par\nobreak

\textbf{Theorem 4.1 (Soundness):} The axiom system $H_{PDL}^{2}$ is sound. That is, for every formula $\varphi$, $\vdash \varphi \Rightarrow \vDash \varphi$.

\textbf{Proof:} We examine each axiom and inference rule in turn.

(1)---(3): Their soundness follows from the soundness of propositional logic.

(4), (5): Since $[\rho]$ and $@$ are normal modal operators, their soundness follows from the soundness of the modal K axiom.

The soundness of (6), (7), and (9) is immediate.

(8), (10): Since the truth value of $@_{(s,t)}\varphi$ does not depend on the point $(u,v)$ at which it is evaluated, (8) and (10) are sound.

(11): By the substitution lemma in Section 2, we have
\[
\begin{aligned}
&M,g,\llbracket s\rrbracket^{M,g},\llbracket t\rrbracket^{M,g}
\vDash\varphi[s/x,t/\lambda]\\
\Longleftrightarrow{}&
M,g[x\mapsto\llbracket s\rrbracket^{M,g},\lambda\mapsto\llbracket t\rrbracket^{M,g}],
\llbracket s\rrbracket^{M,g},\llbracket t\rrbracket^{M,g}\vDash\varphi.
\end{aligned}
\]
The right-hand side is precisely the semantics of $@_{(s,t)}\downarrow(x,\lambda)\varphi$. Hence (11) is sound.

(12), (13): $M,g,w,l\vDash@_{(s,t)}\langle-\rho\rangle u$ if and only if
\[
(\llbracket s\rrbracket^{M,g},\llbracket u\rrbracket^{M,g})
\notin\llbracket\rho\rrbracket^{M,g,\llbracket t\rrbracket^{M,g}},
\]
that is, $M,g,w,l\vDash\neg@_{(s,t)}\langle\rho\rangle u$. Thus (12) is sound; the argument for (13) is analogous.

(14): The left-hand side $M,g,w,l \vDash ((@_{(s,t)}\left\langle \rho \right\rangle u) \land @_{(u,t)}\varphi)$ holds if and only if $(\llbracket s\rrbracket^{M,g},\llbracket u\rrbracket^{M,g}) \in \llbracket \rho\rrbracket ^{M,g,t}$ and $M,g,\llbracket u\rrbracket^{M,g},\llbracket t\rrbracket^{M,g} \vDash \varphi$. The right-hand side $M,g,w,l \vDash @_{(s,t)}\left\langle \rho \right\rangle\varphi$ holds if and only if $\exists a \in W$ such that $(\llbracket s\rrbracket^{M,g},a) \in \llbracket \rho\rrbracket ^{M,g,t}$ and $M,g,a,\llbracket t\rrbracket^{M,g} \vDash \varphi$. It suffices to take $a = \llbracket u\rrbracket^{M,g}$. Hence (14) is sound.

The soundness of (15)---(17) follows from the semantics of the relations.

The soundness of (18)---(21) is immediate.

The soundness of $(MP)$, $(Nec_{\Box})$, and $(Nec@)$ follows from the soundness of propositional logic and normal modal logic.

$(Name)$: Suppose that $\vDash @_{(i,k)}\varphi$. Take arbitrary $M,g,w,l$, and obtain $M'$ by interpreting $i,k$, which do not occur in $\varphi$, as $w,l$, respectively. By the validity of the premise, $M',g,w,l\vDash @_{(i,k)}\varphi$, and hence $M',g,w,l\vDash\varphi$. Since $\varphi$ contains neither $i$ nor $k$, we have $M,g,w,l\vDash\varphi$. Thus the conclusion is valid.

$(BG)$: Suppose that $\vDash @_{(s,t)}\left\langle \rho \right\rangle u \rightarrow @_{(u,t)}\varphi$. Assume, for a contradiction, that $M,g,\llbracket s\rrbracket ,\llbracket t\rrbracket  \nvDash [\rho]\varphi$ for some model and assignment. Then there is $a \in W$ such that $(\llbracket s\rrbracket^{M,g},a) \in \llbracket \rho\rrbracket ^{M,g,t}$ and $M,g,a,\llbracket t\rrbracket  \nvDash \varphi$. Let $M',g'$ be the new model and assignment with $\llbracket u\rrbracket^{M',g'} = a$ and everything else unchanged. Since $u$ occurs in neither $\rho$ nor $\varphi$, and $u \neq s$, we have $(\llbracket s\rrbracket^{M',g'},\llbracket u\rrbracket^{M',g'}) \in \llbracket \rho\rrbracket ^{M',g',t}$ and $M',g',\llbracket u\rrbracket^{M',g'},\llbracket t\rrbracket  \nvDash \varphi$. This contradicts $\vDash @_{(s,t)}\left\langle \rho \right\rangle u \rightarrow @_{(u,t)}\varphi$.

\hfill\(\square\)

We also have the following tautologies:

By (5), (6), and $(Nec@)$, we have

\begin{equation}
@_{(s,t)}(\varphi \land \psi) \leftrightarrow \left( @_{(s,t)}\varphi \land @_{(s,t)}\psi \right)
\tag{22}
\end{equation}

By (8) and (9), we have

\begin{equation}
@_{(s,t)}(u \land v) \rightarrow \left( @_{(s,t)}\varphi \leftrightarrow @_{(u,v)}\varphi \right)
\tag{23}
\end{equation}

We also have the following rule:

By (8), (10), $(Name)$, and $(BG)$, we have

\[\frac{(@_{(s,t)}\left\langle \rho \right\rangle u \land @_{(u,t)}\varphi) \rightarrow \psi}{(@_{(s,t)}\left\langle \rho \right\rangle\varphi) \rightarrow \psi}\qquad\text{(Paste)}\]

where $u$ is a fresh world nominal that does not occur in $\rho,\varphi,\psi,s,t$.

\subsection{Completeness Theorem}

We can now prove that $H_{PDL}^{2}$ is complete. That is, for every set of formulas $\Gamma$ and every formula $\varphi$, $\Gamma \vDash \varphi \Rightarrow \Gamma \vdash \varphi$.

We say that $\Gamma \vdash \varphi$ if and only if there is a finite set of formulas $\Gamma_{0} \subseteq \Gamma$ such that $\vdash (\bigwedge\Gamma_{0}) \rightarrow \varphi$. We say that $\Gamma$ is consistent if and only if $\Gamma \nvdash \bot$.

We say that $\Gamma$ is maximally consistent if and only if $\Gamma$ is consistent and, for every formula $\varphi$, either $\varphi \in \Gamma$ or $\neg\varphi \in \Gamma$.

We say that a consistent set $\Gamma$ is \emph{pasted} if and only if, for every $@_{(s,t)}\left\langle \rho \right\rangle\varphi \in \Gamma$, there is a world nominal $u$ in the language in question such that $@_{(s,t)}\left\langle \rho \right\rangle u \in \Gamma$ and $@_{(u,t)}\varphi \in \Gamma$.

Let $L_{p}'$ denote the extension of $L_{p}$ obtained by taking $Wnom' = Wnom \cup \{ i_{0},i_{1},...\},\ Lnom' = Lnom \cup \{ k_{0},k_{1},...\}$, where $\{ i_{0},i_{1},...\},\{ k_{0},k_{1},...\}$ are countable sets of nominals disjoint from $Wnom,Lnom$. Then $L_{p}'$ is a conservative extension of $L_{p}$. We have the following lemma:

\textbf{Lemma 4.2 (Pasted Lindenbaum Lemma):} For every consistent set $\Gamma$ in $L_{p}$, there is a pasted maximally consistent set $\Sigma$ in the extended language $L_{p}'$ such that $\Gamma \cup \{ i_{0},k_{0}\} \cup \left\{ @_{(i_{x},k_{0})}x \middle| x \in Wvar \right\} \cup \left\{ @_{(i_{0},k_{\lambda})}\lambda \middle| \lambda \in Lvar \right\} \subseteq \Sigma$, where $i_{x},k_{\lambda}$ are pairwise distinct fresh nominals different from $i_{0},k_{0}$.

\textbf{Proof:} First, we show that $\Gamma \cup \{ i_{0},k_{0}\}$ is consistent. Otherwise, there would be a finite set of formulas $\Theta \subseteq \Gamma$ such that $\vdash (i_{0} \land k_{0}) \rightarrow \neg(\bigwedge\Theta)$. By $(Nec@)$ and $(K@)$, we have $\vdash @_{(i_{0},k_{0})}\neg(\bigwedge\Theta)$, and by $(Name)$, $\vdash \neg(\bigwedge\Theta)$, contrary to the consistency of $\Gamma$.

Next, we show that, for every consistent set $\Gamma$ with $\{ i_{0},k_{0}\} \subseteq \Gamma$, the set $\Gamma \cup \{@_{(i_{x},k_{0})}x\}$ is consistent, where $i_x$ is a fresh world nominal that does not occur in the original set. Suppose, for a contradiction, that $\Gamma \cup \{@_{(i_{x},k_{0})}x\}$ is inconsistent. Then there is a finite set of formulas $\Theta \subseteq \Gamma$ such that $\vdash \bigwedge\Theta \rightarrow \neg@_{(i_{x},k_{0})}x$. By $(SubW)$, replace the fresh nominal $i_x$ throughout this theorem by the world variable $x$, obtaining $\vdash \bigwedge\Theta \rightarrow \neg@_{(x,k_{0})}x$. By (7), $\vdash @_{(x,k_{0})}x$, and hence $\vdash \neg(\bigwedge\Theta)$, contrary to the consistency of $\Gamma$. Similarly, applying $(SubL)$ to the fresh lifting-scheme nominal $k_\lambda$ shows that $\Gamma \cup \{@_{(i_{0},k_{\lambda})}\lambda\}$ is consistent.

Enumerate all formulas of $L_{p}'$ as $\theta_{0},\ \theta_{1},...$, and construct sets $\Sigma_{0} \subseteq \Sigma_{1} \subseteq ...$, where $\Sigma_{0}: = \Gamma \cup \{ i_{0},k_{0}\} \cup \left\{ @_{(i_{x},k_{0})}x \middle| x \in Wvar \right\} \cup \left\{ @_{(i_{0},k_{\lambda})}\lambda \middle| \lambda \in Lvar \right\}$. The preceding consistency argument applies each time a fresh naming nominal is added. Since any proof of inconsistency uses only finitely many premises, all the conditions naming variables can be added one by one, so $\Sigma_0$ is consistent. Now suppose that $\Sigma_{n}$ is consistent. If $\Sigma_{n} \cup \{\theta_{n}\}$ is consistent and $\theta_{n}$ is not of the form $@_{(s,t)}\left\langle \rho \right\rangle\varphi$, let $\Sigma_{n + 1}: = \Sigma_{n} \cup \{\theta_{n}\}$. If $\Sigma_{n} \cup \{\theta_{n}\}$ is consistent and $\theta_{n}$ is of the form $@_{(s,t)}\left\langle \rho \right\rangle\varphi$, choose an unused world nominal $u_{n} \in Wnom'$ and let $\Sigma_{n + 1}: = \Sigma_{n} \cup \{\theta_{n},@_{(s,t)}\left\langle \rho \right\rangle u_{n},@_{(u_{n},t)}\varphi\}$. If $\Sigma_{n} \cup \{\theta_{n}\}$ is inconsistent, let $\Sigma_{n+1}:=\Sigma_n\cup\{\neg\theta_n\}$.

If both $\Sigma_{n} \cup \{\theta_{n}\}$ and $\Sigma_{n+1}:=\Sigma_n\cup\{\neg\theta_n\}$ were inconsistent, there would be finite sets of formulas $\Theta_{1},\Theta_{2} \subseteq \Sigma_{n}$ such that $\vdash \neg(\theta_{n} \land (\bigwedge\Theta_{1})), \vdash \neg(\neg\theta_{n} \land (\bigwedge\Theta_{2}))$. Thus $\vdash \neg((\bigwedge\Theta_{1}) \land (\bigwedge\Theta_{2}))$, that is, $\vdash \neg\bigwedge(\Theta_{1}\cup\Theta_{2})$, contrary to the consistency of $\Sigma_{n}$.

We now show that if $\Sigma_{n} \cup \{\theta_{n}\}$ is consistent, then $\Sigma_{n} \cup \{\theta_{n},@_{(s,t)}\left\langle \rho \right\rangle u_{n},@_{(u_{n},t)}\varphi\}$ is consistent. If $\Sigma_{n} \cup \{\theta_{n},@_{(s,t)}\left\langle \rho \right\rangle u_{n},@_{(u_{n},t)}\varphi\}$ were inconsistent, there would be a finite set of formulas $\Theta \subseteq \Sigma_{n}$ such that $\vdash (\theta_{n} \land @_{(s,t)}\left\langle \rho \right\rangle u_{n} \land @_{(u_{n},t)}\varphi) \rightarrow \neg(\bigwedge\Theta)$. By the $(Paste)$ rule, $\vdash (@_{(s,t)}\left\langle \rho \right\rangle\varphi) \rightarrow \neg(\bigwedge\Theta)$, contrary to the consistency of $\Sigma_{n} \cup \{\theta_{n}\}$. Hence $\Sigma_{n} \cup \{\theta_{n},@_{(s,t)}\left\langle \rho \right\rangle u_{n},@_{(u_{n},t)}\varphi\}$ is consistent.

Thus $\Sigma_{n + 1}$ is consistent. By induction, $\Sigma_{n}$ is consistent for every $n\in\mathbb N$. Let $\Sigma = \bigcup_{n\in\mathbb N}\Sigma_{n}$. Since $\theta_{n}$ ranges over all formulas of $L_{p}'$, $\Sigma$ is a pasted maximally consistent set, and $\Gamma \cup \{ i_{0},k_{0}\} \subseteq \Sigma$.

\hfill\(\square\)

For the countable language $L_{p}$, let $\Gamma$ be any pasted maximally consistent set satisfying $\{ i_{0},k_{0}\} \cup \left\{ @_{(i_{x},k_{0})}x \middle| x \in Wvar \right\} \cup \left\{ @_{(i_{0},k_{\lambda})}\lambda \middle| \lambda \in Lvar \right\} \subseteq \Gamma$. Fix $\Gamma,i_{0},k_{0}$. For every $(s,t)$, let $\Delta_{(s,t)}: = \left\{ \varphi \middle| @_{(s,t)}\varphi \in \Gamma \right\}$. We then have:

\textbf{Lemma 4.3:} $\Delta_{(s,t)}$ is a pasted maximally consistent set.

\textbf{Proof:} First, we show that $\Delta_{(s,t)}$ is consistent. Otherwise, there would be a finite set of formulas $\Theta \subseteq \Delta_{(s,t)}$ such that $\vdash \neg(\bigwedge\Theta)$. By $(Nec@)$ and (22), we have $\vdash \neg(\bigwedge\{ @_{(s,t)}\varphi\mid\varphi\in\Theta\})$, contrary to the consistency of $\Gamma$.

Next, we show that $\Delta_{(s,t)}$ is pasted. If $@_{(u,v)}\left\langle \rho \right\rangle\varphi \in \Delta_{(s,t)}$, then $@_{(s,t)}@_{(u,v)}\left\langle \rho \right\rangle\varphi \in \Gamma$. By (8) and $(MP)$, $@_{(u,v)}\left\langle \rho \right\rangle\varphi \in \Gamma$. Since $\Gamma$ is pasted, there exists $a$ such that $@_{(u,v)}\left\langle \rho \right\rangle a \in \Gamma,@_{(a,v)}\varphi \in \Gamma$. By $(Nec@)$, $@_{(s,t)}@_{(u,v)}\left\langle \rho \right\rangle a \in \Gamma,@_{(s,t)}@_{(a,v)}\varphi \in \Gamma$. Hence $@_{(u,v)}\left\langle \rho \right\rangle a \in \Delta_{(s,t)},@_{(a,v)}\varphi \in \Delta_{(s,t)}$, so $\Delta_{(s,t)}$ is pasted.

Finally, we show that $\Delta_{(s,t)}$ is maximally consistent. If $\varphi \notin \Delta_{(s,t)}$, then $@_{(s,t)}\varphi \notin \Gamma$. By the maximal consistency of $\Gamma$, $\neg@_{(s,t)}\varphi \in \Gamma$, that is, $@_{(s,t)}\neg\varphi \in \Gamma$, and thus $\neg\varphi \in \Delta_{(s,t)}$.

\hfill\(\square\)

\textbf{Lemma 4.4:} $@_{(u,v)}\varphi \in \Delta_{(s,t)}$ if and only if $@_{(u,v)}\varphi \in \Gamma$.

\textbf{Proof:} If $@_{(u,v)}\varphi \in \Gamma$, then by (8), $@_{(s,t)}@_{(u,v)}\varphi \in \Gamma$, and hence $@_{(u,v)}\varphi \in \Delta_{(s,t)}$. Conversely, if $@_{(u,v)}\varphi \in \Delta_{(s,t)}$, then $@_{(s,t)}@_{(u,v)}\varphi \in \Gamma$. By (8) and $(MP)$, $@_{(u,v)}\varphi \in \Gamma$. \hfill\(\square\)

\textbf{Lemma 4.5:} If $s \land t \in \Gamma$, then $\Delta_{(s,t)} = \Gamma$.

\textbf{Proof:} For every formula $\varphi \in \Gamma$, since $s \land t \in \Gamma$, (9) gives $@_{(s,t)}\varphi \in \Gamma \Leftrightarrow \varphi \in \Gamma$, that is, $\varphi \in \Delta_{(s,t)} \Leftrightarrow \varphi \in \Gamma$. Hence $\Delta_{(s,t)} = \Gamma$.

Let $\Xi_{s}:=\Delta_{(s,k_{0})}$ and $\Omega_{t}:=\Delta_{(i_{0},t)}$. We have:

\textbf{Theorem 4.6 (Well-definedness):} $\Xi,\Omega$ are well-defined; that is, there is a map ${Map}^{\Gamma}$ such that ${Map}^{\Gamma}(\Xi_{s},\Omega_{t}) = \Delta_{(s,t)}$.

\textbf{Proof:} We need to show that if $\Xi_{s} = \Xi_{u}$ and $\Omega_{t} = \Omega_{v}$, then $\Delta_{(s,t)} = \Delta_{(u,v)}$.

If $\Xi_{s} = \Xi_{u}$, then $u \in \Xi_{s}$, that is, $@_{(s,k_{0})}u \in \Gamma$. By (19), $@_{(s,t)}u \in \Gamma$. Similarly, $@_{(s,t)}v \in \Gamma$. Therefore $@_{(s,t)}(u \land v) \in \Gamma$. By (23), for every formula $\varphi$, $@_{(s,t)}\varphi \in \Gamma \Leftrightarrow @_{(u,v)}\varphi \in \Gamma$. Hence $\Delta_{(s,t)} = \Delta_{(u,v)}$. \hfill\(\square\)

\textbf{Definition 4.7 (Canonical Model):}

Let

\[
\begin{aligned}
W^{\Gamma}&=\left\{\Xi_s\mid s\in Wnom\prime\cup Wvar\right\},\\
{Lf}^{\Gamma}&=\left\{\Omega_t\mid t\in Lnom\prime\cup Lvar\right\},\\
\Xi_sR^{\Gamma}\Xi_u
&\Longleftrightarrow
\text{for every formula }\varphi\text{, if }[r]\varphi\in\Xi_s
\text{, then }\varphi\in\Xi_u,\\
V^{\Gamma}&=\langle V_p^{\Gamma},\eta_W^{\Gamma},\eta_L^{\Gamma}\rangle,\\
V_p^{\Gamma}(p)&=\left\{\Xi_s\mid p\in\Xi_s\right\},\\
\eta_W^{\Gamma}(i)&=\Xi_i,\qquad
\eta_L^{\Gamma}(k)=\Omega_k.
\end{aligned}
\]

The canonical model is then denoted by:

\[M^{\Gamma}: = \left\langle W^{\Gamma},{Lf}^{\Gamma},R^{\Gamma},V^{\Gamma} \right\rangle.\]

For each variable $x \in Wvar,\lambda \in Lvar$, let $i_x$ be a world nominal satisfying $@_{(i_x,k_0)}x\in\Gamma$, and let $k_\lambda$ be a lifting-scheme nominal satisfying $@_{(i_0,k_\lambda)}\lambda\in\Gamma$. If there are $i_{1},i_{2}$ such that $@_{(i_1,k_0)}x\in\Gamma$ and $@_{(i_2,k_0)}x\in\Gamma$, then (5), (7), (8), (9), and $(Nec@)$ give $\Xi_{i_{1}} = \Xi_{i_{2}}$. The case of lifting-scheme variables is analogous. Thus $\Xi_{i_{x}},\Omega_{k_{\lambda}}$ exist and are unique.

The canonical assignment to variables is then:

\[g^{\Gamma} = \left\langle g_{W}^{\Gamma},g_{L}^{\Gamma} \right\rangle,\qquad g_{W}^{\Gamma}(x) = \Xi_{i_{x}},\quad g_{L}^{\Gamma}(\lambda) = \Omega_{k_{\lambda}}.\]

We have the following lemmas:

\textbf{Lemma 4.8 (Relation Lemma):} For every relation term $\rho$, $(\Xi_{s},\Xi_{u}) \in \llbracket \rho\rrbracket ^{M^{\Gamma},g^{\Gamma},\Omega_{t}} \Leftrightarrow \left\langle \rho \right\rangle u \in \Delta_{(s,t)}.$

\textbf{Lemma 4.9 (Truth Lemma):} For every formula $\varphi$, $M^{\Gamma},g^{\Gamma},\Xi_{s},\Omega_{t} \vDash \varphi \Leftrightarrow \varphi \in \Delta_{(s,t)}.$

\textbf{Proof:} Lemmas 4.8 and 4.9 are proved by simultaneous induction on recursive depth.

\textbf{Relation terms:}

When $rk(\rho) = 0$, we have $\rho = r$. We need to show that $\Xi_{s}R^{\Gamma}\Xi_{u} \Leftrightarrow @_{(s,t)}\left\langle r \right\rangle u \in \Gamma$. First, we prove the right-to-left direction. Suppose that $@_{(s,t)}\langle r\rangle u \in \Gamma$. By Axiom 21, we also have $@_{\left( s,k_{0} \right)}\langle r\rangle u \in \Gamma$. Let $\Xi_{s} = \Delta_{s,k_{0}}$, and take any formula $[ r]\chi \in \Xi_{s}$. Then $@_{\left( s,k_{0} \right)}[ r]\chi \in \Gamma$. By (14) and the duality of the box and diamond operators, we obtain $\left( @_{(s,k_0)}\langle r\rangle u\land @_{(s,k_0)}[r]\chi\right)\rightarrow @_{(u,k_0)}\chi$. Thus $\chi \in \Xi_{u}$, and hence $\Xi_{s}R^{\Gamma}\Xi_{u}$. Next, we prove the left-to-right direction. Suppose that $\Xi_{s}R^{\Gamma}\Xi_{u}$, and assume, for a contradiction, that $@_{(s,t)}\langle r\rangle u \notin \Gamma$. By maximal consistency and Axiom 21, $\neg@_{\left( s,k_{0} \right)}\langle r\rangle u \in \Gamma$, that is, $@_{\left( s,k_{0} \right)}[ r]\neg u \in \Gamma$, and thus $[ r]\neg u \in \Xi_{s}$. Since $\Xi_{s}R^{\Gamma}\Xi_{u}$, we have $\neg u \in \Xi_{u}$; however, Axiom 7 and \textbf{K@} give $u \in \Xi_{u}$, a contradiction. Hence $\Xi_{s}R^{\Gamma}\Xi_{u} \Leftrightarrow @_{(s,t)}\left\langle r \right\rangle u \in \Gamma$ holds.

Suppose that, for every formula $\varphi$ and relation $\rho$ of depth $rk < n$, both $(\Xi_{s},\Xi_{u}) \in \llbracket \rho\rrbracket ^{M^{\Gamma},g^{\Gamma},\Omega_{t}} \Leftrightarrow \left\langle \rho \right\rangle u \in \Delta_{(s,t)}$ and $M^{\Gamma},g^{\Gamma},\Xi_{s},\Omega_{t} \vDash \varphi \Leftrightarrow \varphi \in \Delta_{(s,t)}$ hold. Now consider a relation term $\sigma$ with $rk(\rho) = n$.

If $\sigma = - \rho$, then $(\Xi_{s},\Xi_{u}) \in \llbracket  - \rho \rrbracket ^{M^{\Gamma},g^{\Gamma},\Omega_{t}} \Leftrightarrow (\Xi_{s},\Xi_{u}) \notin \llbracket \rho\rrbracket ^{M^{\Gamma},g^{\Gamma},\Omega_{t}}$. By the induction hypothesis, this is equivalent to $\left\langle \rho \right\rangle u \notin \Delta_{(s,t)}$, that is, $\left\langle - \rho \right\rangle u \in \Delta_{(s,t)}$. The cases $\sigma = \rho^{- 1}$ and $\sigma = \rho \cup \tau$ are analogous.

Suppose that $\sigma = \rho;\tau$. First, we prove the left-to-right direction. If $(\Xi_{s},\Xi_{u}) \in \llbracket \rho;\tau\rrbracket ^{M^{\Gamma},g^{\Gamma},\Omega_{t}}$, then there is $X \in W^{\Gamma}$ such that $\left( \Xi_{s},X \right) \in \llbracket \rho\rrbracket ^{M^{\Gamma},g^{\Gamma},\Omega_{t}}$ and $\left( X,\Xi_{u} \right) \in \llbracket \tau\rrbracket ^{M^{\Gamma},g^{\Gamma},\Omega_{t}}$. Let $h \in Wnom\prime \cup Wvar$ be a world term such that $X = \Xi_{h}$. By the induction hypothesis, $@_{(s,t)}\langle\rho\rangle h \in \Gamma,@_{(h,t)}\langle\tau\rangle u \in \Gamma$. By \textbf{Bridge}, $@_{(s,t)}\langle\rho\rangle\langle\tau\rangle u \in \Gamma$, and by \textbf{Seq}, $@_{(s,t)}\langle\rho;\tau\rangle u \in \Gamma$. Next, we prove the right-to-left direction. Suppose that $@_{(s,t)}\langle\rho;\tau\rangle u \in \Gamma$. By \textbf{Seq}, $@_{(s,t)}\langle\rho\rangle\langle\tau\rangle u \in \Gamma$. By the witness property, there is a world nominal $h$ such that $@_{(s,t)}\langle\rho\rangle h \in \Gamma$ and $@_{(h,t)}\langle\tau\rangle u \in \Gamma$. By the induction hypothesis, $\left( \Xi_{s},\Xi_{h} \right) \in \llbracket \rho\rrbracket ^{M^{\Gamma},g^{\Gamma},\Omega_{t}}$ and $\left( \Xi_{h},\Xi_{u} \right) \in \llbracket \tau\rrbracket ^{M^{\Gamma},g^{\Gamma},\Omega_{t}}$. Therefore $\left( \Xi_{s},\Xi_{u} \right) \in \llbracket \rho;\tau\rrbracket ^{M^{\Gamma},g^{\Gamma},\Omega_{t}}$.

Suppose that $\sigma = ?\varphi$. If $(\Xi_{s},\Xi_{u}) \in \llbracket ?\varphi\rrbracket ^{M^{\Gamma},g^{\Gamma},\Omega_{t}}$, this is equivalent to $\Xi_{s} = \Xi_{u}\ $ and $M^{\Gamma},g^{\Gamma},\Xi_{s},\Omega_{t} \vDash \varphi$. By the induction hypothesis, this is equivalent to $u \in \Delta_{s,t}$ and $\varphi \in \Delta_{s,t}$, hence to $(\varphi \land u) \in \Delta_{s,t}$, that is, $\langle?\varphi\rangle u \in \Delta_{s,t}$.

\textbf{Formulas:}

When $rk(\varphi) = 0$, either $\varphi = p$ is an atomic proposition, in which case Axiom 18 gives $M^{\Gamma},g^{\Gamma},\Xi_{s},\Omega_{t} \vDash p \Leftrightarrow \Xi_{s} \in V_{p}^{\Gamma}(p) \Leftrightarrow p \in \Xi_{s} \Leftrightarrow p \in \Delta_{s,t}$, or $\varphi = u \in Wnom \cup Wvar$, in which case $M^{\Gamma},g^{\Gamma},\Xi_{s},\Omega_{t} \vDash u \Leftrightarrow \Xi_{s} = \llbracket u\rrbracket^{M^{\Gamma},g^{\Gamma}} \Leftrightarrow \Xi_{s} = \Xi_{u} \Leftrightarrow u \in \Delta_{s,t}$. The case $\varphi = v \in Lnom \cup Lvar$ is analogous.

Suppose that, for every formula $\varphi$ and relation $\rho$ of depth $rk < n$, both $(\Xi_{s},\Xi_{u}) \in \llbracket \rho\rrbracket ^{M^{\Gamma},g^{\Gamma},\Omega_{t}} \Leftrightarrow \left\langle \rho \right\rangle u \in \Delta_{(s,t)}$ and $M^{\Gamma},g^{\Gamma},\Xi_{s},\Omega_{t} \vDash \varphi \Leftrightarrow \varphi \in \Delta_{(s,t)}$ hold. Now consider a formula $\varphi$ with $rk(\varphi) = n$.

If $\varphi = \neg\psi$, then by the maximal consistency of $\Delta_{s,t}$, $M^{\Gamma},g^{\Gamma},\Xi_{s},\Omega_{t} \vDash \neg\psi \Leftrightarrow M^{\Gamma},g^{\Gamma},\Xi_{s},\Omega_{t} \nvDash \psi$. By the induction hypothesis, this is equivalent to $\psi \notin \Delta_{s,t}$, that is, $\neg\psi \in \Delta_{s,t}$.

If $\varphi = \ \psi \land \chi$, then $M^{\Gamma},g^{\Gamma},\Xi_{s},\Omega_{t} \vDash \psi \land \chi \Leftrightarrow \; M^{\Gamma},g^{\Gamma},\Xi_{s},\Omega_{t} \vDash \psi\ \text{and}\ M^{\Gamma},g^{\Gamma},\Xi_{s},\Omega_{t} \vDash \chi$. By the induction hypothesis, this is equivalent to $\;\psi \in \Delta_{s,t}\ \text{and}\ \chi \in \Delta_{s,t}$, that is, $\;\psi \land \chi \in \Delta_{s,t}$.

If $\varphi = @_{(u,v)}\psi$, then $M^{\Gamma},g^{\Gamma},\Xi_{s},\Omega_{t} \vDash @_{(u,v)}\psi$ is equivalent to $M^{\Gamma},g^{\Gamma},\Xi_{u},\Omega_{v} \vDash \psi$. By the induction hypothesis, $\psi \in \Delta_{u,v}$, that is, $@_{(u,v)}\psi \in \Gamma$, and thus $@_{(u,v)}\psi \in \Delta_{s,t}$.

For $\varphi = \downarrow (x,\lambda)\psi$, $M^{\Gamma},g^{\Gamma},\Xi_{s},\Omega_{t} \vDash \downarrow (x,\lambda)\psi \Leftrightarrow \; M^{\Gamma},g^{\Gamma}\left[ x \mapsto \Xi_{s},\lambda \mapsto \Omega_{t} \right],\Xi_{s},\Omega_{t} \vDash \psi$, that is, $M^{\Gamma},g^{\Gamma},\Xi_{s},\Omega_{t} \vDash \psi[ s/x,t/\lambda]$. Since $rk\left( \psi[ s/x,t/\lambda] \right) = rk(\psi)$, the induction hypothesis gives $\;\psi[ s/x,t/\lambda] \in \Delta_{s,t}$, that is, $\; \downarrow (x,\lambda)\psi \in \Delta_{s,t}$.

Suppose that $\varphi = [\rho]\psi$. First, we prove the right-to-left direction. If $[\rho]\psi \in \Delta_{s,t}$, take any $X \in W^{\Gamma}$ satisfying $\left( \Xi_{s},X \right) \in \llbracket \rho\rrbracket ^{M^{\Gamma},g^{\Gamma},\Omega_{t}}$, and let $u \in Wnom\prime \cup Wvar$ be such that $X = \Xi_{u}$. By the induction hypothesis, $@_{(s,t)}\langle\rho\rangle u \in \Gamma$. Moreover, $[\rho]\psi \in \Delta_{s,t}$ is equivalent to $@_{(s,t)}[\rho]\psi \in \Gamma$. By (14) and modal duality, $@_{(u,t)}\psi\in\Gamma$, that is, $\psi \in \Delta_{u,t}$. By the induction hypothesis, $M^{\Gamma},g^{\Gamma},\Xi_{u},\Omega_{t} \vDash \psi$. Thus every $\rho$-successor of $s$ satisfies $\psi$, so $M^{\Gamma},g^{\Gamma},\Xi_{s},\Omega_{t} \vDash [\rho]\psi$. Next, we prove the left-to-right direction. Suppose that $M^{\Gamma},g^{\Gamma},\Xi_{s},\Omega_{t} \vDash [\rho]\psi$, and assume, for a contradiction, that $[\rho]\psi \notin \Delta_{s,t}$. Then $\langle\rho\rangle\neg\psi \in \Delta_{s,t}$, that is, $@_{(s,t)}\langle\rho\rangle\neg\psi \in \Gamma$. By the witness property, there is a world nominal $u$ such that $@_{(s,t)}\langle\rho\rangle u \in \Gamma$ and $@_{(u,t)}\neg\psi \in \Gamma$. The induction hypothesis therefore gives $\left( \Xi_{s},\Xi_{u} \right) \in \llbracket \rho\rrbracket ^{M^{\Gamma},g^{\Gamma},\Omega_{t}}$ and $M^{\Gamma},g^{\Gamma},\Xi_{u},\Omega_{t} \vDash \neg\psi$, contrary to $M^{\Gamma},g^{\Gamma},\Xi_{s},\Omega_{t} \vDash [\rho]\psi$. \qed

\textbf{Theorem 4.10 (Completeness):} For every set of formulas $\Sigma$ and every formula $\varphi$, $\Sigma \vDash \varphi \Rightarrow \Sigma \vdash \varphi$.

\textbf{Proof:} Suppose, for a contradiction, that $\Sigma \nvdash \varphi$. Then $\Sigma \cup \{\neg\varphi\}$ is consistent. By Lemma 4.2, there is a pasted maximally consistent set $\Gamma$ such that $\Sigma \cup \{\neg\varphi,i_{0},k_{0}\} \subseteq \Gamma$. Since $\{ i_{0},k_{0}\} \subseteq \Gamma$, that is, $i_{0} \land k_{0} \in \Gamma$, Lemma 4.5 gives $\Gamma = \Delta_{(i_{0},k_{0})}$. Hence $M^{\Gamma},g^{\Gamma},\Xi_{i_{0}},\Omega_{k_{0}} \vDash \Gamma$, that is, $M^{\Gamma},g^{\Gamma},\Xi_{i_{0}},\Omega_{k_{0}} \vDash \Sigma$ and $M^{\Gamma},g^{\Gamma},\Xi_{i_{0}},\Omega_{k_{0}} \vDash \neg\varphi$. This contradicts $\Sigma \vDash \varphi$.

\hfill\(\square\)

\textbf{Corollary 4.11:} The axiom system \(H_{PDL}^{2}\)+\(\left( @_{(s,t)}\langle r\rangle u \land @_{(u,t)}\langle r\rangle h \right) \rightarrow @_{(s,t)}\langle r\rangle h\)+\(@_{(s,t)}\langle r\rangle s\) is sound and strongly complete with respect to all frames whose underlying relation is a preorder.

\textbf{Proof:} The two axioms express reflexivity and transitivity, respectively. The canonical countermodel constructed above therefore belongs to the class of preorder frames. \hfill\(\square\)

\section{Conclusion}

Starting from an example of the framing effect, this paper treats the lifting scheme as an independent dimension of evaluation and develops a preference logic with dynamically changing lifting schemes. We give its syntax, semantics, and axiom system, and prove the soundness and strong completeness of the axiom system. We prove that dynamic changes in lifting schemes cannot be reduced to dynamic changes in the preference relation, thereby establishing the value of treating lifting schemes as an independent dimension and studying their dynamics. In addition, regarding order properties, we prove that there is no uniform decision procedure for the preservation of reflexivity or transitivity by singleton-conservative first-order liftings. For common preference lifting schemes, we provide a characterization of preorder preservation through a sufficient and natural condition.

Future research may investigate the interaction between changes in preference lifting schemes and changes in preferences. Since preference lifting schemes determine the success conditions for preference dynamics, the logical properties relating different lifting schemes to preference dynamics remain to be explored. The dynamics of preference lifting may also be placed in the context of social dynamics to study dynamic changes in the preference logic of communities.


\begin{thebibliography}{99}
\bibitem{ref1} Tversky A, Kahneman D. The Framing of Decisions and the Psychology of Choice{[}J{]}. Science, 1981, 211(4481): 453--458.
\bibitem{ref2} Li S, Xie X. A New Look at the ``Asian Disease'' Problem: A Choice between the Best Possible Outcomes or between the Worst Possible Outcomes?{[}J{]}. Thinking \& Reasoning, 2006, 12(2): 129--143.
\bibitem{ref3} Hansson S O. Preference Logic{[}M{]}//Gabbay D M, Guenthner F, eds.~Handbook of Philosophical Logic. 2nd ed., Vol. 4. Dordrecht: Kluwer, 2001: 319--393.
\bibitem{ref4} Liu F. Reasoning about Preference Dynamics{[}M{]}. Synthese Library, Vol. 354. Dordrecht: Springer, 2011.
\bibitem{ref5} van Benthem J, Liu F. Dynamic Logic of Preference Upgrade{[}J{]}. Journal of Applied Non-Classical Logics, 2007, 17(2): 157--182.
\bibitem{ref6} Barberà S, Bossert W, Pattanaik P K. Ranking Sets of Objects{[}M{]}//Barberà S, Hammond P J, Seidl C, eds.~Handbook of Utility Theory. Vol. 2. Boston: Kluwer, 2004: 893--977.
\bibitem{ref7} Kannai Y, Peleg B. A Note on the Extension of an Order on a Set to the Power Set{[}J{]}. Journal of Economic Theory, 1984, 32(1): 172--175.
\bibitem{ref8} Barberà S, Pattanaik P K. Extending an Order on a Set to the Power Set: Some Remarks on Kannai and Peleg's Approach{[}J{]}. Journal of Economic Theory, 1984, 32(1): 185--191.
\bibitem{ref9} Bossert W. On the Extension of Preferences over a Set to the Power Set: An Axiomatic Characterization of a Quasi-Ordering{[}J{]}. Journal of Economic Theory, 1989, 49(1): 84--92.
\bibitem{ref10} Bossert W, Pattanaik P K, Xu Y. Choice under Complete Uncertainty: Axiomatic Characterizations of Some Decision Rules{[}J{]}. Economic Theory, 2000, 16(2): 295--312.
\bibitem{ref11} Arlegi R. A Note on Bossert, Pattanaik and Xu's ``Choice under Complete Uncertainty: Axiomatic Characterizations of Some Decision Rules''{[}J{]}. Economic Theory, 2003, 22(1): 219--225.
\bibitem{ref12} Pattanaik P K, Peleg B. An Axiomatic Characterization of the Lexicographic Maximin Extension of an Ordering over a Set to the Power Set{[}J{]}. Social Choice and Welfare, 1984, 1: 113--122.
\bibitem{ref13} Nitzan S, Pattanaik P K. Median-Based Extensions of an Ordering over a Set to the Power Set: An Axiomatic Characterization{[}J{]}. Journal of Economic Theory, 1984, 34: 252--261.
\bibitem{ref14} Kreps D M. A Representation Theorem for ``Preference for Flexibility''{[}J{]}. Econometrica, 1979, 47(3): 565--577.
\bibitem{ref15} Pattanaik P K, Xu Y. On Ranking Opportunity Sets in Terms of Freedom of Choice{[}J{]}. Recherches Économiques de Louvain, 1990, 56: 383--390.
\bibitem{ref16} Bossert W, Pattanaik P K, Xu Y. Ranking Opportunity Sets: An Axiomatic Approach{[}J{]}. Journal of Economic Theory, 1994, 63(2): 326--345.
\bibitem{ref17} Plotkin G D. A Powerdomain Construction{[}J{]}. SIAM Journal on Computing, 1976, 5(3): 452--487.
\bibitem{ref18} Smyth M B. Power Domains{[}J{]}. Journal of Computer and System Sciences, 1978, 16(1): 23--36.
\bibitem{ref19} Brewka G, Truszczyński M, Woltran S. Representing Preferences among Sets{[}C{]}//Proceedings of the Twenty-Fourth AAAI Conference on Artificial Intelligence. AAAI Press, 2010: 273--278.
\bibitem{ref20} Shi C, Sun Y. Logic of Convex Order{[}J{]}. Studia Logica, 2021, 109(5): 1019--1047.
\bibitem{ref21} Brandt F, Brill M. Necessary and Sufficient Conditions for the Strategyproofness of Irresolute Social Choice Functions{[}C{]}//Proceedings of the 13th Conference on Theoretical Aspects of Rationality and Knowledge. ACM, 2011. DOI: 10.1145/2000378.2000394.
\bibitem{ref22} Leal J F, Pino Pérez R. A Weak Version of Barberà--Kelly's Theorem{[}J{]}. Revista Colombiana de Matemáticas, 2017, 51(2): 173--194.
\bibitem{ref23} Maly J, Wallner J P. Ranking Sets of Defeasible Elements in Preferential Approaches to Structured Argumentation: Postulates, Relations, and Characterizations{[}C{]}//Proceedings of the AAAI Conference on Artificial Intelligence. 2021, 35(7): 6435--6443.
\bibitem{ref24} Halpern J Y. Defining Relative Likelihood in Partially-Ordered Preferential Structures{[}J{]}. Journal of Artificial Intelligence Research, 1997, 7: 1--24.
\bibitem{ref25} Dubois D, Fargier H. A Unified Framework for Order-of-Magnitude Confidence Relations{[}C{]}//Proceedings of the Twentieth Conference on Uncertainty in Artificial Intelligence. AUAI Press, 2004: 138--145.
\bibitem{ref26} Harrison-Trainor M, Holliday W H, Icard T F. Preferential Structures for Comparative Probabilistic Reasoning{[}C{]}//Proceedings of the AAAI Conference on Artificial Intelligence. 2017: 1135--1141. Revised version: arXiv:2104.02287, 2021.
\bibitem{ref27} Darmann A, Klamler C. Using the Borda Rule for Ranking Sets of Objects{[}J{]}. Social Choice and Welfare, 2019, 53: 399--414.
\bibitem{ref28} Fu X, Zhao Z. A Logic for Preference Lifting under Uncertainty and Its Decidability{[}C{]}//Logic and Argumentation. Lecture Notes in Computer Science, Vol. 14156. Cham: Springer, 2023: 213--223.
\bibitem{ref29} van Benthem J, Girard P, Roy O. Everything Else Being Equal: A Modal Logic Approach to Ceteris Paribus Preferences{[}J{]}. Journal of Philosophical Logic, 2009, 38: 83--125.
\bibitem{ref30} Andréka H, Ryan M, Schobbens P Y. Operators and Laws for Combining Preference Relations{[}J{]}. Journal of Logic and Computation, 2002, 12(1): 13--53.
\bibitem{ref31} van Benthem J. Relational Patterns, Partiality, and Set Lifting in Modal Semantics{[}M{]}//Weiss Y, Birman R, eds.~Saul Kripke on Modal Logic. Outstanding Contributions to Logic, Vol. 30. Cham: Springer, 2024: 93--119.
\bibitem{ref32} Goncharov S, Hofmann D, Nora P, Schröder L, Wild P. Kantorovich Functors and Characteristic Logics for Behavioural Distances{[}C{]}//Foundations of Software Science and Computation Structures. Cham: Springer, 2023: 46--67. DOI: 10.1007/978-3-031-30829-1\_3.
\bibitem{ref33} Goncharov S, Hofmann D, Nora P, Schröder L, Wild P. A Point-Free Perspective on Lax Extensions and Predicate Liftings{[}J{]}. Mathematical Structures in Computer Science, 2024, 34(1): 98--127.
\bibitem{ref34} Gargov G, Passy S, Tinchev T. Modal Environment for Boolean Speculations{[}M{]}//Skordev D G, ed.~Mathematical Logic and Its Applications. New York: Plenum Press, 1987: 253--263.
\bibitem{ref35} Blackburn P, Tzakova M. Hybrid Completeness{[}J{]}. Logic Journal of the IGPL, 1998, 6(4): 625--650.
\bibitem{ref36} Areces C, ten Cate B. Hybrid Logics{[}M{]}//Blackburn P, van Benthem J, Wolter F, eds.~Handbook of Modal Logic. Amsterdam: Elsevier, 2006: 821--868.
\bibitem{ref37} Sano K, Tojo S. Dynamic Epistemic Logic for Channel-Based Agent Communication{[}C{]}//Logic and Its Applications. Lecture Notes in Computer Science, Vol. 7750. Berlin: Springer, 2013: 109--120.
\bibitem{ref38} Lavrov I A. Effective Inseparability of the Set of Identically True and Finitely Refutable Formulas of Some Elementary Theories{[}J{]}. Algebra i Logika, 1963, 2(1): 5--18. 
\bibitem{ref39} Maly J. Ranking Sets of Objects: How to Deal with Impossibility Results{[}D{]}. Vienna: Technische Universität Wien, 2020.
\end{thebibliography}
\end{document}